\documentclass[showpacs,twocolumn,prl,aps]{revtex4}
\usepackage{graphicx}
\begin{document}

\title{Variational study of the one dimensional $t-J$ model}
\author{Hong-Yu Yang}
\affiliation{Center for Advanced Study, Tsinghua University,
Beijing, 100084, P. R. China}
\author{Tao Li}
\affiliation{Department of Physics, Renmin University of China,
Beijing, 100872, P. R. China}
\date{{\small \today}}

\begin{abstract}
We find the Gutzwiller projected Fermi sea wave function(GWF) has
the correct phase structure to describe the kink nature of the doped
holes in the ground state of the one dimensional $t-J$ model. We
find the failure of the GWF for general value of $J/t$ and electron
density $n$ can be attributed to the residual charge correlation in
the ground state. We find such residual charge correlation is well
described by a XXZ-type effective Hamiltonian. Based on these
observations, a Pfaffian-type variational wave function is proposed
and is found to reproduce correctly the global phase diagram and
corresponding correlation functions of the one dimensional $t-J$
model, including the Luther-Emery phase in the low electron density
and large $J/t$ region.\end{abstract}

\maketitle

\section{I.\hspace{0.5cm}  Introduction}
The Gutzwiller projected wave functions(GWFs) are widely used to
approximate the ground state of the $t-J$ model and the Heisenberg
model. In these models, local electronic correlation, as manifested
in the no double occupancy constraint of electrons, plays a vital
role in determining the low energy physics. Such strong local
correlation make these systems difficult to study analytically. In
the variational approach based on GWFs, these models are first
treated in the mean field approximation in which the local
constraint is relaxed to a global one. Then the local constraint is
enforced afterwards by the Gutzwiller projection which simply
filters out the unphysical components with doubly occupied sites in
the mean field state.

The above variational strategy is used extensively in the study of
the high temperature superconductors and quantum antiferromagnets.
After many years of efforts, it is now believed that the Gutzwiller
projected d-wave BCS state describe well the superconducting state
of the high temperature
superconductors\cite{yokoyama1,gros,paramekanti1,paramekanti2}.
Quite recently, progress is also made on the understanding of the
quasiparticle properties above such a
state\cite{yunoki1,yunoki2,randeria,nave}. The same kind of wave
function is also used in recent studies on the exotic orders and
exotic excitations of frustrated quantum
antiferromagnet\cite{ivanov,yunoki3,paramekanti3,sorella}.

An unresolved issue about the GWF is that it is not clear if the
posteriorly executed projection can capture the kinematic effect of
the local constraints, even qualitatively. In this paper, we address
this issue with the one dimensional $t-J$ model.

The one dimensional $t-J$ model has been studied extensively by a
broad band of methods including Bethe-Ansatz solution, conformal
field theories\cite{bares,kawakami,kuramoto}, quantum Monte
carlo\cite{assaad}, exact diagonalization\cite{ogata1}, and also
Variational Monte Carlo
calculations\cite{ogata2,gebhard,hellberg,chen1,yokoyama2,kobayashi,chen2}.
Many properties concerning the ground state of this model are now
well established. This give us the unique opportunity to judge the
validity of a given approximation. The one dimensional $t-J$ model
is exactly soluble at $J/t=0$ and $J/t=2$. For $J/t=0$, the spin and
the charge degree of freedom of the system are totally
separated\cite{ogata2}. The spin part is described by the Heisenberg
model on the squeezed chain with doped holes removed, while the
charge part is described by a noninteracting spinless Fermion
system. For $J/t=2$, the system is supersymmetric and it is found
that the GWF provides a fairly accurate approximation for the ground
state of the system\cite{yokoyama2,kuramoto}. For general value of
$J/t$ and electron density $n$, the system is a Tomonaga-Luttinger
liquid(TLL) below a critical value $J_{c}/t$ around 2.5. The
correlation exponent of the TLL varies continuesly with $J/t$ and
$n$\cite{ogata1}. For $J/t>J_{c}/t$, the system is unstable toward
phase separation. For small $n$ and $J/t>2$, there is also a small
region in which the system exhibit a spin gap\cite{chen1,chen2}.

The Gutzwiller projected Fermi sea wave function and its variants
has long been used to describe the ground state of the one
dimensional $t-J$ model. It is well known that this wave function
provides an excellent description of the undoped case of the model,
namely the spin $\frac{1}{2}$ Heisenberg spin chain\cite{gebhard}.
However, the same wave function is not that satisfactory for the
doped system, except for the supersymmetric case of $J/t=2$. For
example, it fails to predict the TLL behavior in the small $J/t$
region. A $2\mathrm{k}_{F}$ peak in the spin structure factor is
also missed by this wave function. Since the wave function is
parameter free, it also gives no clue on the origin of the spin gap
state and the phase separation at large $J/t$.

It is generally believed that the problems with the GWF originate
from the insufficient account of the charge correlation in the
system. Along this line of thinking, various kind of Jastrow factor
are proposed to remedy the drawbacks of GWF. For example, Hellberg
and Mele introduced a long range Jastrow factor of the form
$|F(r_{i\uparrow},r_{j\downarrow})|^{\nu} $ and succeed in
reproducing the TLL behavior, where
$|F(r_{i\uparrow},r_{j\downarrow})|^{\nu} $ is a Slater determinant
of all the electron positions\cite{hellberg}. Yokoyama and Ogata
found a short range repulsive Jastrow factor is able to restore the
$2\mathrm{k}_{F}$ peak in the spin structure factor, while a
sufficiently attractive Jastrow factor can cause phase
separation\cite{yokoyama2}. However, both wave functions have
difficulties in reproducing the correct phase diagram. For example,
the spin gap state is missed in both wave functions. At the same
time, both wave functions predict a fully phase-separated state
along the boundary of phase separation, which is in fact an
oversimplification\cite{ogata1}. More importantly, no understanding
on the physical origin of the proposed Jastrow factor is available
and it is hard to judge if a similar modification is relevant for
higher dimensional system.

For the sake of possible extension to higher dimensional case, it is
important to know the reason that the simple GWF fails before any
modification on it is made. As mentioned above, it is the residual
charge correlation in the system which is responsible for the
failure of GWF. In this paper we make this statement more precise by
showing that the GWF has the correct phase structure to describe the
kink nature of the doped holes in the ground state of the one
dimensional $t-J$ model. In fact, we find the spin structure factor
of the GWF in the squeezed chain coordinate is almost identical to
that of a half filled spin chain. Thus the missing $2\mathrm{k}_{F}$
peak in the spin structure factor for small $J/t$ should be
recovered if the removed holes are reinserted into the squeezed
chain in the right manner.

The physical origin of the residual charge correlation can be easily
seen if one reformulate the GWF in terms of the slave Boson
theory\cite{kotliar,lee}. In the slave Boson theory, the constrained
electron operator is decomposed as
$\hat{c}^{\dagger}_{i,\sigma}=f^{\dagger}_{i,\sigma}b_{i}$, in which
$f^{\dagger}_{i,\sigma}$ is a spin $\frac{1}{2}$ neutral Fermion
called spinon and $b_{i}$ is a spinless charge 1 Boson called holon.
The local constraint now takes the form of an equality,
$\sum_{\sigma}f^{\dagger}_{i,\sigma}f_{i,\sigma}+b^{\dagger}_{i}b_{i}=1$.
In terms of the slave Boson theory, the GWF corresponds to a state
with all holon condensed into the zero momentum state. However, the
holon is not a true Boson but a hard core Boson as a result of the
local constraint. For general value of $J/t$, there is also an
effective attraction between the holons caused by the exchange term
of the $t-J$ model. Thus a XXZ-type effective Hamiltonian should be
a good approximation for the residual charge correlation.

Based on these observations, a Pfaffian-type variational wave
function is proposed for the ground state of the one dimensional
$t-J$ model. This wave function, which has only one parameter,
reproduce well the global phase diagram of the model, including the
Luther-Emery(LE) phase in the small $n$ and large $J/t$ region. It
is found that this wave function also reproduces well various
correlation functions of the system and provides a refined picture
for the phase separation at large $J/t$.

The paper is organized as follows. Section II is devoted to the
investigation of the properties of the GWF. In Section III, the new
variational scheme and the Pfaffian-type wave function are
introduced. The phase diagram and correlation functions determined
from this new variational wave function are presented in Section IV.
Section V summarize the paper and includes a discussion on related
issues.

\section{II.\hspace{0.5cm}  The GWF}
The one dimensional $t-J$ model reads
\begin{equation}\label{1}
    \mathcal{H}=-t\sum_{i,\sigma}(\hat{c}_{i\sigma}^{\dagger}\hat{c}_{i+1,\sigma}+h.c.)
    +J\sum_{i}(\mathbf{S}_{i} \cdot
    \mathbf{S}_{i+1}-\frac{1}{4}n_{i}n_{i+1}),
\end{equation}
in which
$\mathbf{S}_{i}=\frac{1}{2}\sum_{\alpha\beta}\hat{c}_{i\alpha}^{\dagger}
\mathbf{\sigma}_{\alpha\beta}\hat{c}_{i\beta}$ and
$n_{i}=\sum_{\alpha}\hat{c}_{i\alpha}^{\dagger}\hat{c}_{i\alpha}$.
The electron in this model is subjected to the constraint of no
double occupancy
\begin{equation}\label{2}
    \sum_{\alpha}\hat{c}_{i\alpha}^{\dagger}\hat{c}_{i\alpha}\leq1.
\end{equation}

The ground state of the one dimensional $t-J$ model is governed by a
well defined phase structure. This can be most easily seen at half
filling when the system reduces to the Heisenberg spin chain. For
the Heisenberg model, it is well known that the ground state satisfy
the Marshall sign rule\cite{marshall,weng}. The rule says that the
ground state wave function is real in the Ising basis and its sign
is given by $(-1)^{N_{\downarrow}}$ up to a global phase, where
$N_{\downarrow}$ denotes the number of down spins in the even
sublattice. This sign rule is a manifestation of the
antiferromagnetic spin correlation in the ground state. With such a
sign rule, one easily verify that
$\langle\mathbf{S}_{i}\cdot\mathbf{S}_{j}\rangle\leq0$ for $i$ and
$j$ belonging to different sublattices and
$\langle\mathbf{S}_{i}\cdot\mathbf{S}_{j}\rangle\geq0$ for $i$ and
$j$ belonging to the same sublattice.

The ground state at finite doping is governed by a similar sign
rule. It can be easily checked that all matrix elements of the $t-J$
Hamiltonian are negative definite in a wave function that satisfy
the Marshall sign rule on the squeezed chain. The squeezed chain is
the chain in which the sites occupied by the doped holes are
removed. This can be seen by noting that the motion of holes in this
model do not disturb the spin configuration on the squeezed chain.
Thus, the ground state of the one dimensional $t-J$ model should
satisfy such a modified Marshall sign rule. With such a modified
Marshall sign rule, one easily see that the holes in the ground
state behaves as an antiphase domain wall for spin.

Now we show that the GWF satisfies the Marshall sign rule on the
squeezed chain. The GWF reads
\begin{equation}\label{4}
    |\mathrm{GWF}\rangle=\prod_{i}(1-n_{i\uparrow}n_{i\downarrow})|\mathrm{FS}\rangle,
\end{equation}
in which $|\mathrm{FS}\rangle$ denotes the simple Fermi sea. In the
natural basis
$\prod_{i,j}c_{i\uparrow}^{\dagger}c_{j\downarrow}^{\dagger}|0\rangle$,
the amplitude of GWF is given by the following Vandermont
determinant
\begin{equation}\label{8}
    \Psi(\{i\},\{j\})=\psi_{PW}\prod_{\alpha<\beta}(Z_{i_{\alpha}}-Z_{i_{\beta}})\prod_{l<m}(Z_{j_{l}}-Z_{j_{m}}),
\end{equation}
in which
$\psi_{PW}=\exp[-i\mathrm{k}_{F}(\sum_{\alpha}i_{\alpha}+\sum_{l}j_{l})]$
is a plane wave factor, $Z_{i_{\alpha}}=\exp(i\frac{2\pi
i_{\alpha}}{N})$ and $Z_{j_{l}}=\exp(i\frac{2\pi j_{l}}{N})$ are
chord coordinates of up spins and down spins.

Now we exchange the up spin at site $i_{1}$ and the down spin at
site $j_{1}$. The resultant change in phase is given by
\begin{equation}
    \Delta\Phi=\arg(\prod_{\alpha>1}\frac{Z_{i_{\alpha}}-Z_{j_{1}}}{Z_{i_{\alpha}}-Z_{i_{1}}}\prod_{l>1}
    \frac{Z_{j_{l}}-Z_{i_{1}}}{Z_{j_{l}}-Z_{j_{1}}}).
\end{equation}
Since $|Z|=1$,
$\arg(\frac{Z_{i_{\alpha}}-Z_{j_{1}}}{Z_{i_{\alpha}}-Z_{i_{1}}})$ is
nothing but the angle in the segment $Z_{i_{1}}-Z_{j_{1}}$ in the
unit circle. Noting the fact that in a circle the angles in the same
segment equal one another and sum of the opposite angles of
quadrilaterals equals $\pi$, one easily find that
$\Delta\Phi=N_{c}\pi$, in which $N_{c}$ denotes the number of
electrons between site $i_{1}$ and site $j_{1}$. Taking into account
the sign due to Fermion exchange, one find the change in phase is in
accordance with the modified Marshall sign rule. Following
essentially the same steps, one can also verify the case of
exchanging a hole and an electron.

\begin{figure}[h!]
\includegraphics[width=6.5cm,angle=0]{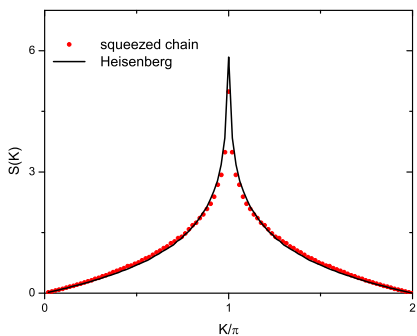}
\caption{Spin structure factor of the GWF in the squeezed coordinate
as compared with that of a half filled spin chain. The calculation
is done on a 204 sites lattice which is quarter filled.}
 \label{fig1}
\end{figure}

Thus the GWF has the right phase structure to describe the ground
state of the one dimensional $t-J$ model and the kink nature of the
doped holes in it. In fact, this conclusion can be made even
stronger. In Figure 1, we plot the spin structure factor of the GWF
in the squeezed coordinate and compare it with that of a half filled
spin chain. We see the two are almost identical with each other.
Since the spin degree of freedom is described exactly by a
Heisenberg model on the squeezed chain at $J/t=0$, while the GWF
provides an exceedingly good approximation for $J/t=2$, it is
natural to expect that the same behavior to hold for arbitrary $J/t$
and $n$.

Two conclusions follow directly from the above reasoning. First,
since the spin correlation on the squeezed chain is already well
described by the GWF, the missing $2\mathrm{k}_{F}$ peak in the spin
structure factor should be recovered if the removed holes are
correctly reinserted into the squeezed chain, or, the the missing
$2\mathrm{k}_{F}$ peak should be attributed to the residual charge
correlation in the system. Second, since the squeezed spin chain
picture is argued to hold for arbitrary $J/t$ and $n$, a single wave
function may suffice to describe the whole phase diagram of the one
dimensional $t-J$ model, including the spin gap phase at small $n$
and large $J/t$.

\section{III.\hspace{0.5cm}  The new variational scheme}
The origin of the residual charge correlation can be most easily
seen by reformulating the GWF in terms of the slave Boson theory. In
the slave Boson theory, the constrained electron operator is
decomposed as
$\hat{c}_{i,\sigma}^{\dagger}=f_{i,\sigma}^{\dagger}b_{i}$, in which
$f_{i,\sigma}^{\dagger}$ represents the Fermionic spinon and $b_{i}$
represents the Bosonic holon. In terms of these slave particles, the
$t-J$ model reads
\begin{eqnarray*}
  \mathcal{H} &=& \mathcal{H}_{t}+\mathcal{H}_{J} \\
  \\
  \mathcal{H}_{t} &=& -t\sum_{i,\sigma}(f_{i,\sigma}^{\dagger}f_{i+1,\sigma}b_{i+1}^{\dagger}b_{i}+h.c.) \\
  \mathcal{H}_{J} &=& \frac{J}{2}
  \sum_{i}b_{i}b_{i}^{\dagger}b_{i+1}b_{i+1}^{\dagger}(\mathbf{S}_{i}^{f} \cdot
  \mathbf{S}_{i+1}^{f}-\frac{1}{4}n_{i}^{f}n_{i+1}^{f}),
\end{eqnarray*}
in which
$\mathbf{S}_{i}^{f}=\frac{1}{2}\sum_{\alpha\beta}f_{i\alpha}^{\dagger}
\mathbf{\sigma}_{\alpha\beta}f_{i\beta}$ and
$n_{i}^{f}=\sum_{\alpha}f_{i\alpha}^{\dagger}f_{i\alpha}$. The no
double occupancy constraint now takes the form of an equality
\begin{equation}\label{3}
    \sum_{\alpha}f_{i\alpha}^{\dagger}f_{i\alpha}+b_{i}^{\dagger}b_{i}=1.
\end{equation}
When the local constraints Eq.(6) is exactly satisfied, the factor
$b_{i}b_{i}^{\dagger}b_{i+1}b_{i+1}^{\dagger}$ appearing in
$\mathcal{H}_{J}$ plays no role and can be neglected.

In the mean field treatment, an RVB order parameter
$\chi=\sum_{\alpha} <f_{i+1\alpha}^{\dagger}f_{i\alpha}>$ is
introduced to decompose the interaction term. At the same time, the
local constraint is relaxed to a global one. The mean field
Hamiltonian for the spinon and the holon part read\cite{lee}
\begin{eqnarray*}
  \mathcal{H}^{f} &=&  -(tx+\frac{3J\chi}{8})\sum_{i\sigma}(f_{i,\sigma}^{\dagger}f_{i+1\sigma}+h.c.)\\
  \mathcal{H}^{b} &=&  -t\chi\sum_{i}(b_{i}^{\dagger}b_{i+1}+h.c.),
\end{eqnarray*}
in which $x$ is the hole density. The mean field ground state is
given by the product of the spinon Fermi sea and the holon Bose
condensate
\begin{equation}\label{5}
    |\Phi\rangle=(b_{\mathrm{k}=0}^{\dagger})^{N_{h}}
    \prod_{\mathrm{k}\leq\mathrm{k}_{F}}f_{\mathrm{k}\uparrow}^{\dagger}
    f_{\mathrm{k}\downarrow}^{\dagger}|0\rangle.
\end{equation}
When this state is projected into the subspace that satisfy the
constraint Eq.(6), we get the GWF.

In the mean field theory, the holon is a free Boson and condenses in
the ground state. However, due to the local constraint, the holon is
actually a hard core Boson which can not condense in one spatial
dimension. The uncondensed nature of the hard core Boson in 1d
originates from the kinematic effect of the local constraint. Due to
this constraint, the Hilbert space for the one dimensional hard core
Boson system becomes disconnected at the single particle level. We
note for comparison that the Hilbert space of the spinon part is
still connected even when the local constraint is enforced. Thus the
holon should be treated as hard core Boson rather than free Boson.

Another source of the residual charge correlation is provided by the
superexchange term of the $t-J$ model. When two electrons are next
to each other, they enjoy an attraction due to the superexchange.
This attraction is not captured by the mean field order parameter
$\chi$ and should be reintroduced.

Combining these considerations, the residual charge correlation
beyond the GWF should be described by the following XXZ-type
effective Hamiltonian
\begin{equation}\label{8}
    \mathcal{H}_{v}=-\sum_{i}(\hat{b}_{i}^{\dagger}\hat{b}_{i+1}+h.c.)
    -v\sum_{i}\hat{b}_{i}^{\dagger}\hat{b}_{i+1}^{\dagger}\hat{b}_{i+1}\hat{b}_{i},
\end{equation}
in which $\hat{b}_{i}^{\dagger}$ is the operator for hard core Boson
and $v$ is the rescaled attraction. If we denote the ground state of
$\mathcal{H}_{v}$ as $\Lambda_{v}$, then
$\mathrm{P_{G}}\Lambda_{v}|\mathrm{GWF}\rangle$ should be a good
variational wave function for the one dimensional $t-J$ model.

Although $\mathcal{H}_{v}$ is exactly soluble\cite{yangcn}, an
explicit form for $\Lambda_{v}$ is available only in limited cases.
For $v=0$, $\Lambda_{v}$ is nothing but the Hellberg-Mele Jastrow
factor with $\nu=1$. For $v=1$, $\Lambda_{v}$ is a constant and our
proposed wave function reduce to the GWF. At the quarter filling,
$\mathcal{H}_{v}$ exhibit particle-hole symmetry. In a separated
paper\cite{yang}, we show a Hellberg-Mele-type variational wave
function provides an exceedingly good description for the ground
state of the XXZ model in the $S^{z}=0$ sector. However, away from
the particle-hole symmetric point, the Hellberg-Mele wave function
cease to be a good approximation.

For general value of $v$ and Boson density, we have to resort to
approximation. Through the Jordan-Wigner transformation, the XXZ
Hamiltonian can be rewritten as
\begin{equation}\label{8}
    \mathcal{H}_{v}=-\sum_{i}(c_{i}^{\dagger}c_{i+1}+h.c.)
    -v\sum_{i}c_{i}^{\dagger}c_{i+1}^{\dagger}c_{i+1}c_{i},
\end{equation}
in which $c_{i}^{\dagger}$ is a spinless Fermion. For this
Hamiltonian, we adopt the BCS approximation to decouple the
interaction term. The BCS ground state for the spinless Fermion
reads
\begin{equation}\label{9}
    \prod_{\mathrm{k}>0}(u_{\mathrm{k}}+v_{\mathrm{k}}c_{\mathrm{k}}^{\dagger}c_{\mathrm{-k}}^{\dagger})|0\rangle,
\end{equation}
in which
$\frac{v_{\mathrm{k}}}{u_{\mathrm{k}}}=\frac{\Delta_{\mathrm{k}}}{\epsilon_{\mathrm{k}}+\mathrm{E}_{\mathrm{k}}}$,
$\Delta_{\mathrm{k}}=\Delta\sin(\mathrm{k})$,
$\epsilon_{\mathrm{k}}=-2\cos(\mathrm{k})-\mu$ and
$\mathrm{E}_{\mathrm{k}}=\sqrt{\epsilon_{\mathrm{k}}^{2}+\Delta_{\mathrm{k}}^{2}}$.
Here $\Delta$ is the BCS gap for the spinless Fermion and is treated
as the only variational parameter in our theory(the chemical
potential $\mu$ can be determined by the density equation and is not
an independent parameter). In real space, the BCS state for the
spinless Fermion takes the form of a Pfaffian. A Pfaffian is a
square root of the determinant of a antisymmetric matrix of even
order\cite{yunoki4}. In our case, the matrix element of the
antisymmetric matrix is given by
\begin{equation}\label{10}
    \mathrm{f}_{i,j}=\sum_{\mathrm{k}>0}\frac{v_{\mathrm{k}}}{u_{\mathrm{k}}}\sin(\mathrm{k(i-j)}),
\end{equation}
in which $i$ and $j$ denote the coordinates of the spinless
Fermions. Thus our variational wave function for the one dimensional
$t-J$ model is given by
\begin{equation}\label{11}
    \Psi=\mathrm{Pf}(\Delta)|\mathrm{GWF}\rangle,
\end{equation}
in which $\mathrm{Pf}(\Delta)$ is the Pfaffian for the holes which
are now spinless Fermions.

\section{IV.\hspace{0.5cm}  Results}
\subsection{A.  Ground state phase diagram}
The ground state phase diagram determined from the Pfaffian-type
wave function is presented in Figure 2.

\begin{figure}[h!]
\includegraphics[width=6.5cm,angle=0]{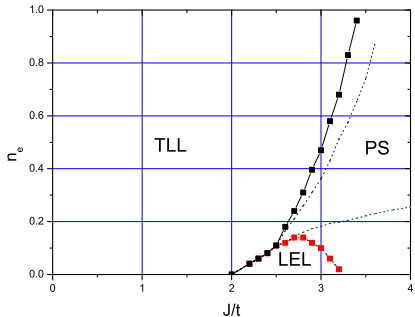}
\caption{Ground state phase diagram of the one dimensional $t-J$
model determined from the Pfaffian-type variational wave function.
The dotted lines indicate the boundaries for the existence of
locally stable phases. Here TLL denotes Tomonaga-Luttinger liquid,
LEL denotes Luther-Emery liquid, while PS denotes phase separated
state.}
 \label{fig2}
\end{figure}

The phase diagram contains three distinct phases. For small and
intermediate value of J/t, the system is in the TLL phase in which
both charge and spin excitation are gapless. For larger value of
J/t, the system is unstable towards phase separation. At small $n$
and large J/t, there is a small region in which the system exhibits
a spin gap. In the spin gap phase, the charge excitation is still
gapless. Following the convention, this phase is termed Luther-Emery
liquid.

The phase boundaries are determined as follows. To illustrate the
idea, we plot the variational energy as a function of the electron
density for $J/t=2$ and $J/t=3$ in Figure 3. For $J/t=2$, the energy
curve is concave everywhere so that a homogenous phase is globally
stable for all electron density. For $J/t=3$, a convex region
appears at intermediate values of electron density in the energy
curve. In this case, the boundaries for the globally stable phases
are given by the two tangency points shown in the figure, while the
boundaries for the locally stable phases are given by the two
inflexion points.

\begin{figure}[h!]
\includegraphics[width=6.5cm,angle=0]{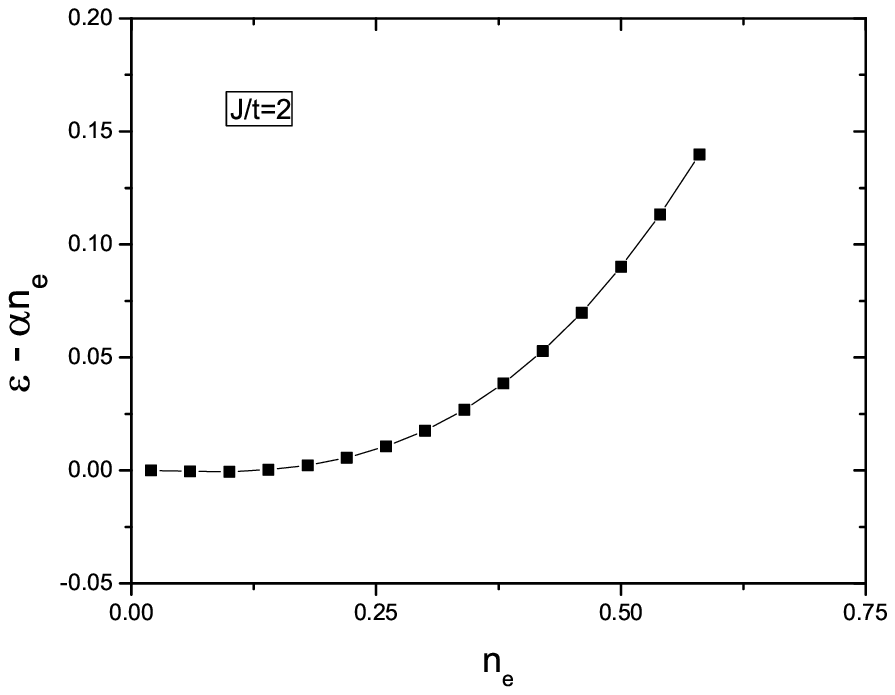}
\includegraphics[width=6.5cm,angle=0]{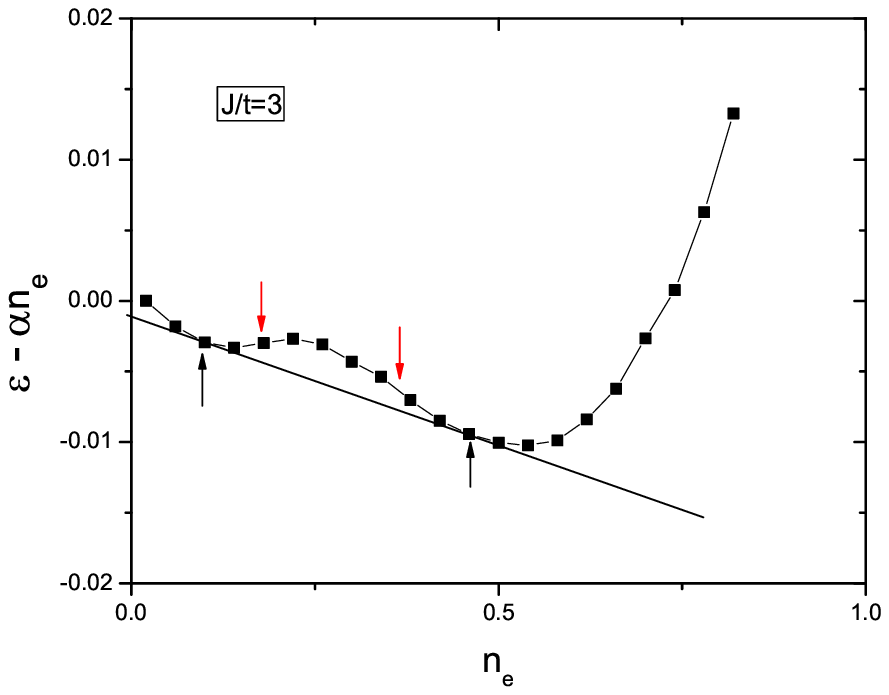}
\caption{Variational energy per site $\epsilon$ as a function of the
electron density $\mathrm{n_{e}}$ for $J/t=2$ and $J/t=3$. For
clarity's sake, a linear decreasing background of the energy is
subtracted.
$\alpha=\frac{\mathrm{d}\varepsilon}{\mathrm{d}\mathrm{n_{e}}}|_{\mathrm{n_{e}}\rightarrow0}$
is the initial slope of the energy curve. The arrows above the curve
indicate the locations of the inflexion points, while the arrows
below the curve indicate the locations of the tangency points. The
phase boundaries are determined from these points as explained in
the text.}
 \label{fig3}
\end{figure}

For electron density that lies between the two tangency points, the
system is unstable towards phase separation. The density of the
phase separated phases are given by two tangency points. For
$2.5<J/t<3.2$, the system phase separates into a hole rich phase and
an electron rich phase. For $3.2<J/t<3.4$, the hole rich phase is
replaced by a empty phase. For $J/t>3.4$, a fully phase separated
state is realized in which the electron rich phase is replaced by a
half filled spin chain.

The convex region of the energy curve diminishes to zero at about
$J/t=2.5$. The phase boundary between the TLL phase and the LEL
phase for $J/t<2.5$ is determined by examining the infrared behavior
of the spin structure factor $S(q)$. In the spin gap phase, $S(q)$
should be quadratic at small $q$, while in the TLL phase a linear
behavior is expected\cite{hohenberg}. For the charge excitation, a
similar criteria exists on the density structure factor $N(q)$.

The existence of the LEL phase is quite unexpected from the point of
view of the mean field theory. In the mean field theory, the spinon
is still described by a filled Fermi sea which is by definition
gapless. However, after Gutzwiller projection the spinon get
entangled with the holon. Such an entanglement change drastically
the spin correlation of the system.

\subsection{B.  Correlation functions}
Four correlation functions are evaluated in this work. They are the
momentum distribution function defined as
\begin{equation}\label{12}
    n(k)=\frac{1}{2N}\sum_{i,j,\sigma}\langle c_{i\sigma}^{\dagger}c_{j\sigma} \rangle
    e^{ik(r_{i}-r_{j})},
\end{equation}
the spin structure factor defined as
\begin{equation}\label{12}
    S(k)=\frac{4}{N}\sum_{i,j}\langle S_{i}^{z}S_{j}^{z} \rangle
    e^{ik(r_{i}-r_{j})},
\end{equation}
the charge structure factor defined as
\begin{equation}\label{12}
    C(k)=\frac{1}{N}\sum_{i,j}(\langle n_{i}n_{j} \rangle- \langle n_{i} \rangle \langle n_{j} \rangle)
    e^{ik(r_{i}-r_{j})},
\end{equation}
and the pair correlation function defined as
\begin{equation}\label{12}
    P(k)=\frac{1}{N}\sum_{i,j}(\langle \Delta_{i}^{\dagger}\Delta_{j} \rangle
    e^{ik(r_{i}-r_{j})},
\end{equation}
in which $\Delta_{i}$ is the annihilation operator for a
nearest-neighboring pair
\begin{equation}\label{12}
    \Delta_{i}=\frac{1}{\sqrt{2}}(c_{i\uparrow}c_{i+1\downarrow}-c_{i\downarrow}c_{i+1\uparrow}).
\end{equation}

First we present the result for the TLL phase. The correlation
functions for $J/t=0,1$ and 2 at quarter filling are shown in Figure
4. For comparison's sake, we also plot the result calculated from
the Hellberg-Mele wave function. From the figure we see that the
correlation functions calculated from the Pfaffian-type wave
function are almost identical with that calculated from the
Hellberg-Mele wave function, apart from the small deviations due to
critical fluctuations. Since the Pfaffian is derived from a BCS mean
field approximation in which a gap opens up, the residual charge
correlation described by it is short ranged. Thus the Pfaffian-type
wave function should exhibit Fermi-liquid behavior, as is clear in
Figure 4. To recover the critical fluctuations, one should go beyond
the mean field approximation.
\begin{figure}[h!]
\includegraphics[width=9cm,angle=0]{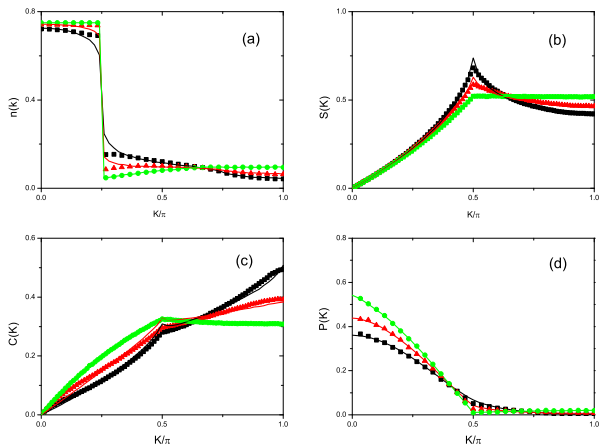}
\caption{(a) The momentum distribution function $n(k)$, (b) the spin
structure factor $S(k)$, (c) the charge structure factor $C(k)$, and
(d) the singlet pairing correlation function $P(k)$ at quarter
filling for J/t=0(black square), J/t=1(red up triangle), and
J/t=2(green circle). The solid lines denote the result calculated
from the Hellberg-Mele variational wave function.}
 \label{fig4}
\end{figure}

For $J/t>2$, the Pfaffian-type wave function becomes less
satisfactory for the quarter filled system. In Figure 5, we plot the
correlation functions of the quarter filled system at $J/t=2.5$ and
3, the latter of which is very close to the boundary of phase
separation. Near the boundary of phase separation, the Hellberg-Mele
wave function starts to develop charge instability, as is clear from
Fig. 5(c). This tendency is missed by the Pfaffian-type wave
function. Instead, a structure at $2k_{F}$ remains evident in the
correlation functions. This is to be expected, since we start from a
Fermionic description of the residual charge correlation. In fact,
it is quite amazing that the BCS approximation remains to be a good
approximation for $J/t$ as high as 2.5(the optimized value for the
BCS gap $\Delta$ is approximately given by $J/t$ at quarter
filling).

\begin{figure}[h!]
\includegraphics[width=9cm,angle=0]{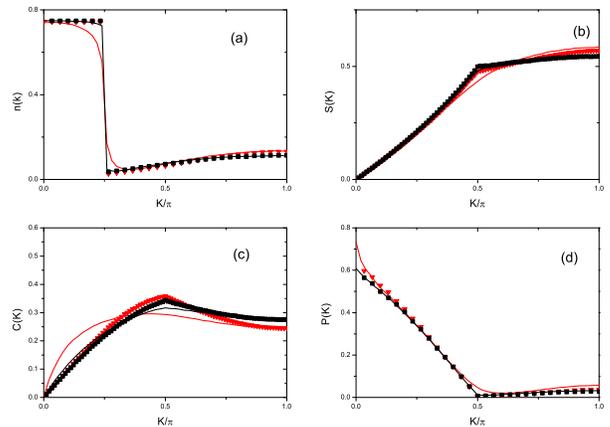}
\caption{(a) The momentum distribution function $n(k)$, (b) the spin
structure factor $S(k)$, (c) the charge structure factor $C(k)$, and
(d) the singlet pairing correlation function $P(k)$ at quarter
filling for J/t=2.5(black square) and J/t=3(red up triangle). The
solid lines denote the result calculated from the Hellberg-Mele
variational wave function.}
 \label{fig5}
\end{figure}

To quantify the above discussion, we plot in Figure 6 the relative
error in the variational energy for both the Pfaffian-type wave
function and the Hellberg-Mele wave function. For small $J/t$, the
energy of the Pfaffian-type wave function is slightly lower than
that of the Hellberg-Mele wave function. For larger value of $J/t$,
the ordering is reversed. However, both wave functions give good
estimate for the ground state energy before phase separation.
\begin{figure}[h!]
\includegraphics[width=9cm,angle=0]{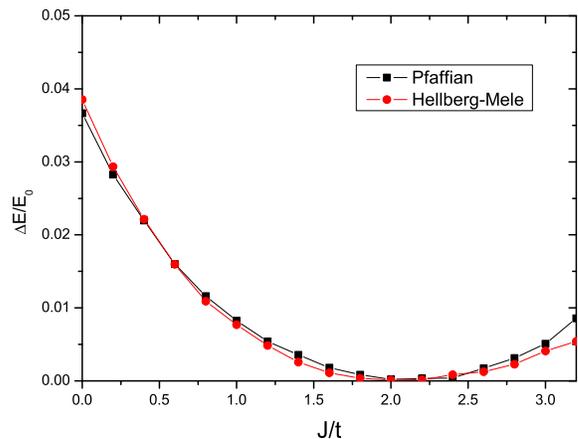}
\caption{Relative error in variational energy at quarter filling.
The exact value of the ground state energy is taken from
\cite{yokoyama2}.}
 \label{fig6}
\end{figure}

Although the Hellberg-Mele wave function provides a good description
for the quarter filled system, it fails badly at low electron
density. On the other hand, the Pfaffian-type wave function
describes quite well the physics in the low density regime,
including the spin gap phase at large $J/t$. To illustrate this, we
plot in Figure 7 the error in variational binding energy for a
single pair of electrons calculated from both wave functions. From
the figure we see the Pfaffian-type wave function is almost exact
for all values of $J/t$ in the low density limit. We think this
explains why the spin gap phase can be correctly reproduced by the
Pfaffian-type variational wave function.
\begin{figure}[h!]
\includegraphics[width=9cm,angle=0]{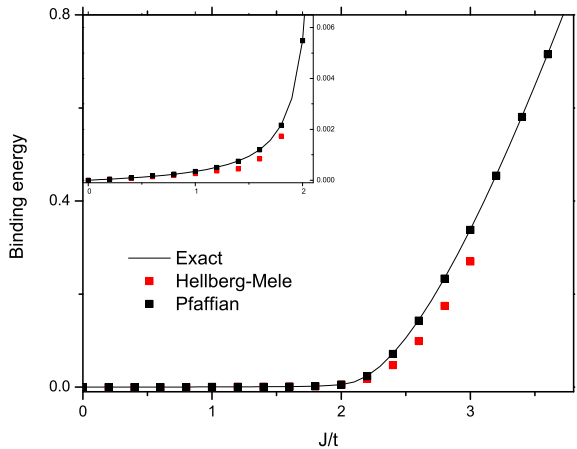}
\caption{Error in variational binding energy for a single pair of
electrons calculated from the Pfaffian-type wave function and the
Hellberg-Mele wave function. The inset shows an expanded view of the
$0<J/t<2$ region.}
 \label{fig7}
\end{figure}

Now we present the correlation functions for the LEL phase at small
$n$ and large $J/t$. In Figure 8, the correlation functions for
$J/t=2.8$ and $n=0.06$, a system deeply inside the LEL phase, are
plotted. As mentioned above, the spin gap manifests itself in the
quadratic behavior of the spin structure factor in the small $q$
limit. We note that spin gap state is metastable in a much larger
region than that of the LEL phase.
\begin{figure}[h!]
\includegraphics[width=9cm,angle=0]{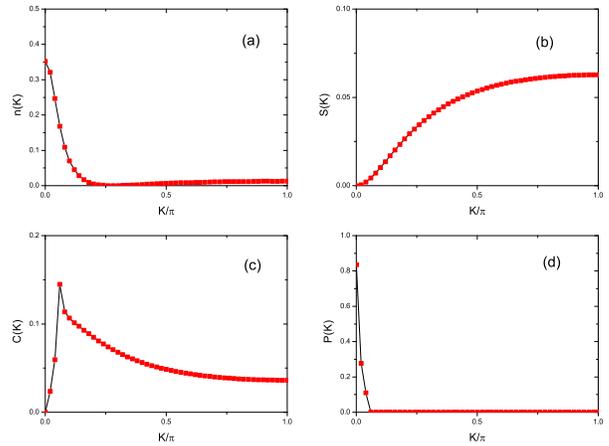}
\caption{Correlation functions for $J/t=2.8$ and $n=0.06$, a system
deeply inside the spin gap phase.(a) The momentum distribution
function $n(k)$, (b) the spin structure factor $S(k)$, (c) the
charge structure factor $C(k)$, and (d) the singlet pairing
correlation function $P(k)$.}
 \label{fig8}
\end{figure}

\section{V.\hspace{0.5cm}  Summary and Discussion}
In this paper, we have carried out a variational study of the one
dimensional $t-J$ model. We find the failure of the simple GWF
should be attributed to the residual charge correlation.
Reformulating the GWF in terms of the slave Boson theory, we find
the residual charge correlation should be described by a XXZ-type
effective Hamiltonian. Based on this observations, a Pfaffian-type
variational wave function is proposed for the one dimensional $t-J$
model. We find this wave function, which has only one variational
parameter, reproduces correctly the global phase diagram and the
corresponding correlation functions.

It is interesting to note the way in which the spin correlation is
affected by the charge degree of freedom in this model. Through the
investigation of the phase structure of the ground state wave
function, we find the doped holes behave as anti-phase domain walls
for the spin correlation. We show further the spin degree of freedom
of the system is well approximated by a half filled spin chain in
the squeezed coordinates throughout the phase diagram. For small
electron density, the effect of the charge degree of freedom on the
spin part can be so drastic as to induce a gap in the excitation
spectrum of the latter. This spin gap is beyond mean field
description and should be attributed to the strong entanglement of
spin and charge degree of freedom in the projected subspace.

It is also interesting to note the effect of the local constraint
for this system. In conventional GWF, the effect of the local
constraint is taken into account posteriorly by filtering out the
unphysical components in the unprojected state. In this paper, we
find this procedure may fail when the kinematic effect of the local
constraint is essential for establishing(or, more accurately,
destroying) the mean field correlation in the unprojected state. The
one dimensional $t-J$ model provides a particular example of this
type. In the one dimensional $t-J$ model, the Hilbert space for the
charge degree of freedom is disconnected at the single particle
level due to the local constraint. When the local constraint is
relaxed, the connectivity of the Hilbert space for the charge degree
of freedom is changed in a qualitative manner. Such a change in the
connectivity of the Hilbert space is responsible for the appearance
of the Bose condensation of the charged particle in the mead field
theory and is ultimately responsible for the failure of the GWF to
describe the Tomonaga-Luttinger behavior of the system.

For a full understanding of the residual charge correlation in the
one dimensional $t-J$ model, one should also take into account the
attraction due to the exchange term. In the mean field treatment,
the exchange term is decoupled in the
$f_{i\sigma}^{\dagger}f_{j\sigma}$ channel which can not account for
such a charge correlation effect. We find this attraction
counteracts with the effect of the local constraint and cancels it
out around $J/t=2$. This explains the excellentness of the GWF at
the supersymmetric point.

The critical behavior of the one dimensional $t-J$ model is not
correctly described by the Pfaffian-type wave function. This is
natural since the Pfaffian itself is derived from a BCS-type mean
field treatment. Although single particle condensation is gone in
this treatment, a condensate of pair of spinless Fermion still
exists. To recovery the correct critical behavior, one must get rid
of such a condensate. One way to achieve this is to introduce a
second charge correlator of the Hellberg-Mele-type, as is done in
\cite{chen2} and \cite{kobayashi}. According to our analysis, the
problem of finding a good variational description for the one
dimensional $t-J$ model reduces to that for the much simpler one
dimensional XXZ model. We think the correct critical behavior should
be recovered by a more accurate guess for the ground state wave
function of the latter model.

Finally, we mention possible generalization of the idea used in this
work to the study of the two dimensional $t-J$ model. In two spatial
dimension, the kinematic effect of the local constraint should be
less dramatic since the connectivity of the Hilbert space is not
affected by the local constraint. This can also be seen from the
fact that the two dimensional XXZ model does undergo Bose
condensation at zero temperature. However, it is much subtler to
analyze the interplay between the spin and charge degree of freedom
in two spatial dimension since the two frustrate each other. Thus,
the validity of the simple GWF in two dimension remains to be seen.

This work is supported by NSFC Grant No.90303009.

\end{document}